# VOYAGER 1 AND 2 OBSERVATIONS OF COSMIC RAY INTENSITIES IN THE NORTH-SOUTH HELIOSHEATHS - IMPLICATIONS FOR THE LATITUDE EXTENT OF THE HELIOSPHERIC CURRENT SHEET AND RADIAL STRUCTURE IN THE HELIOSHEATH

**W.R. Webber[1], D.S. Intriligator[2]**

1. New Mexico State University, Department of Astronomy, Las Cruces, NM  88003, USA
2. Carmel Research Center, Space Plasma Laboratory, Santa Monica, CA  90406, USA

# ABSTRACT

The paper describes striking differences in the intensity as a function of radial distance of anomalous and galactic cosmic rays in the N-S heliosheaths as observed by the Voyager 1 and 2 spacecraft respectively.   The radial profiles in the N heliosheath beyond the heliospheric termination shock (HTS) observed at 94 AU by V1 are much more uniform.   The anomalous cosmic ray (ACR) intensities above a few MeV in the N heliosheath reach a maximum at ~2009.5, about 16 AU beyond the heliospheric current sheet (HTS), thus indicating a possible source location for these higher energy particles.   The galactic cosmic ray electron and nuclei intensities continue to increase rapidly throughout the entire N-heliosheath out to 121.7 AU. Two sudden increases of these electrons at 16.6 and 22.2 AU beyond the HTS suggest that the V1 spacecraft passed significant heliosheath structures at these times.   Later, at 121.7 AU, which is a distance of 27.6 AU beyond the initial HTS crossing distance, V1 in the N heliosheath appears to pass beyond the ACR trapping region as the anomalous particles disappear suddenly and completely.   At V2 in the S heliosheath, large 42-21 day periodic intensity variations were observed for a few MeV particles just beyond the HTS crossing distance at 83.7 AU and again about a year later when V2 was ~4 AU beyond this shock.   These are the dominant features of the energetic particle populations in the inner part of the S-heliosheath.   The ACR and electrons later decreased to almost background levels until a sudden increase at 2011.25, at a distance of about 10 AU beyond the HTS.   The features observed by V2 in the S heliosheath relative to those at V1 indicate a much more turbulent and structured S heliosheath in which the latitude of V2 relative to the HCS maximum latitude (sector zone) may play an important role.  Estimates of a S heliosheath trapping/modulation boundary for energetic particles in the range 105-110 AU are obtained in this paper from a comparison with those at 121.7 AU observed in the N hemisphere.

**Introduction**

Voyager 1 (V1) crossed the heliospheric termination shock (HTS) at about 2004.95, at a distance ~94 AU and a N latitude ~33° (Stone, et al., 2005). Since that time this spacecraft, at the end of 2013 at 127 AU, has reached a point 33 AU beyond the original HTS crossing distance. During this 9.0 year time period the intensities of GCR nuclei and electrons increased steadily at V1 so that by the end of 2012 the intensity of >70 MeV nuclei was ~2.5 times that at the end of 2004 and the 6-14 MeV electron intensity increased by a factor ~100 as both of these intensities increase toward the local interstellar intensity. In fact, the more sensitive low rigidity electrons at V1 exhibit many distinct intensity changes between about 2005.0 and 2008.5 which have amplitudes between 10-50% and which are in coincidence with outward propagating solar disturbances and magneto-sonic waves generated at the HTS (Webber, et al., 2009; Washimi, et al., 2012). After about 2008.5 (107 AU) when the intensity-time profiles become even more uniform, two sudden increases in electron intensity, ~20-30%, are observed at V1. These increases and the associated radial intensity gradient changes of both electrons and nuclei appear to be caused by magnetic structures in the outer heliosheath (Burlaga, et al., 2011; Webber, et al., 2012).

At V1 the termination shock particle (TSP) intensities at all energies between 0.5 and about 10 MeV also increased more or less continuously after the HTS crossing, rapidly at first, then more slowly up to about 2009.5, 16 AU beyond the HTS. However, the large continuous intensity increase that is seen for 5-12 MeV GCR electrons is not observed for these lower energy particles that are believed to be accelerated somewhere in the heliosheath. In fact, after 2009.5 the TSP intensities slowly decrease. This decrease amounts to 50% by about 2012.5.

For the ACR, as reflected in the 27-42 MeV/nuc He nuclei, the intensity at V1 also reaches a maximum in 2009, but the intensity changes are very muted and amount to a factor of only 2-3 over the entire heliosheath.

Both the TSP and ACR disappear suddenly and almost completely starting at 2012.65 (Stone, et al., 2013; Krimigis, et al., 2013). We take this location to be the outer trapping boundary of these particles in the heliosheath in the North at this time, which gives a nominal reference thickness of this region to be 27.6 AU (121.7-94.1).

At V2 in the S heliosheath the intensity vs. distance/time profiles of all components are totally different than at V1. For example, after the V2 HTS crossing at ~84 AU at a S latitude of 25° at 2007.66 (Stone, et al., 2008), the GCR 5-12 MeV electron intensity remains nearly constant for the 1st 10 AU beyond the HTS, except for two spikes of intensity observed at about 2007.9 and again at 2009.2 (~1 and 5 AU beyond the HTS) that may represent locally accelerated electrons (McDonald, et al., 2009). A rapid increase of GCR electron intensity begins near the end of 2010, over three years after the HTS crossing, but after about 2011.5 the GCR electron intensity decreases more or less continuously by almost a factor of 2 up to the end of 2012 when V2 is 16.5 AU beyond the HTS crossing distance. This temporal pattern of intensities beyond the HTS is followed by GCR H and He nuclei, but the changes are much more modest.

The TSP at V2 initially have very large 42 and 21 day intensity variations after the HTS crossing with a maximum intensity at ~2007.9 (Webber, et al., 2012). At this time the TSP intensities are ~3 times larger at 1 MeV than at any time beyond the HTS at V1. A second intensity peak occurs at 2009.15 when V2 was ~4.5 AU beyond the HTS. The TSP intensities then slowly decrease over the next 2.5 years to the end of 2012 when their intensity is actually lower than at any time since V2 crossed the HTS.

The overall increase of ACR intensity beyond the HTS at V2 is less than a factor of two until about 8-9 AU beyond the shock crossing distance. Then the ACR intensity increases rapidly by a factor ~10. The ACR decrease slowly over the next 2.0 years to the end of 2012. Their intensity at this time is still ~10 times larger than it was when V2 crossed the HTS, whereas for TSP the intensity is actually lower at the end of 2012 than at the time of the HTS crossing as noted in the previous paragraph.

It is these differences in the intensity-profiles for all three components, GCR, TSP and ACR in the N and S heliosheaths at V1 and V2 that we wish to describe and discuss in this paper along with estimates of the trapping/modulation boundary locations observed and soon to be observed in both the N-S heliosheaths, respectively.

**The N-S Radial Profiles at V1 and V2**

Figure 1 shows the intensity-time profiles of 5-12 MeV electrons at V1 and V2 in the N-S heliosheaths. The data are 5 day running average intensities. The radial distance (in AU) at V2 is indicated along the upper X axis and the time at V2 along the lower X axis. The time at V1 is plus 2.70 years.

Figure 2 is identical to Figure 1 except it is for 130-242 MeV GCR H nuclei. The intensities here are weekly averages.

Figure 3 is identical to Figure 2 except that the intensities are for 28-42.5 MeV/nuc He nuclei which consist mainly of ACR.

Figure 4 is identical to Figure 2 except that the intensities are for 1.9-2.7 MeV H nuclei which consist mainly of TSP.

Each figure is similar in that the V1 and V2 times are synchronized at the time of the V1 and V2 crossings of the HTS. This means delaying the V1 time by 2.70 years. Note that V1 is moving outward at the rate of 3.62 AU/year, whereas for V2 this rate is 3.14 AU/year. The ratio of the outward speeds is therefore 1.145. This is almost the same as the ratio of the HTS, N-S asymmetry of 1.125 that has been observed (e.g., Opher, 2006). This means that if the heliosheath radial dimensions beyond the HTS scale according to the HTS distances, e.g., the various features in each heliosheath including the HP will occur at almost the same "time" in each hemisphere in each figure. And since the "nominal" N heliosheath thickness is 27.6 AU the "nominal" S heliosheath thickness would be expected to be 27.6/1.125=24.5 AU in this scenario. If, however, corresponding "features", including the HP, are indeed observed by V1 and V2 in the N-S heliosheaths, and if they occur earlier in the S heliosheath, then the S heliosheath thickness is compressed relative to the "nominal" thicknesses, or visa-versa.

**Discussion of Data**

**1). GCR electrons at V1 and V2.** For these particles the intensities after the HTS crossing at both V1 and V2 are expected to increase toward their estimated intensities in the local interstellar medium (LIS) which are ~100 times those observed at the time of the HTS crossing (e.g., Webber and Higbie, 2009). The electron intensity at V1 does indeed increase rapidly beyond the HTS and at 121 AU (27.6 AU beyond the HTS crossing distance) the intensity is over 100 times the intensity at the HTS crossing ~7.5 years earlier. This corresponds to an average radial intensity gradient over this time period ~50%/AU (McDonald, et al., 2012). The sudden intensity jumps of electrons at V1 at +16.6 and +22.2 AU beyond the HTS and a new series of electron intensity increases starting at +26.2 AU beyond the HTS (2012.35) seen in Figure 1 indicate significant large scale structures in the N hemisphere outer heliosheath (Webber, et al., 2012; Burlaga and Ness, 2012; Webber and McDonald, 2013).

At V2 the intensity-distance profile of the 5-12 MeV electrons in Figure 1 is totally different than that just described for V1. The electron intensity remains essentially constant for nearly three years at a level near to that inside the HTS except for the sudden intensity increases centered at 2007.9 and 2009.1. Then late in 2010 when V2 is 9 AU beyond the HTS, the intensity increases by a factor ~10 over a distance of less than 2 AU. After ~2011.8 the electron intensity slowly decreased, reaching an intensity level at the end of 2012 (17 AU beyond the HTS crossing distance) which is ~10 times less than that observed at 2012.0 by V1 which is about 20 AU further out and just inside the HP.

The sudden increases for electrons at V2 with maxima at 2007.9 and 2009.15 in Figure 2 are periods when freshly accelerated electrons are believed to be present (McDonald, et al., 2009; Webber, et al., 2012). The 1$^{st}$ electron increase at ~2007.9, just after V2 crossed the HTS, is in a very disturbed period when large periodic variations, ~42 days peak to peak, were observed for both TSP nuclei and electrons (Webber, et al., 2012). The maxima (and minima) of the intensity peaks of these + and – charge particles are ~180° out of phase during these variations which, along with the changing TSP spectrum below a few MeV at this time as described by Cummings, et al., 2010, indicate great turbulence and possible local acceleration of both types of particles.

The 2$^{nd}$ sudden electron increase at V2 at 2009.15 was also accompanied by large 52 day periodic increases of TSP. After about 2009.5 the TSP intensities exhibited a strong 27 day periodicity, lasting for several months. These changes were not reported in the lower energy (KeV) TSP observed by LECP (e.g., Decker, et al., 2010) This periodicity was not evident for electrons. This 2$^{nd}$ time period is discussed in a later section of this paper.

**2). GCR 133-242 MeV H nuclei at V1 and V2 (Figure 2)**. At V1 the intensity time profile of these GCR shows a steady increase from the time of the HTS crossing in late 2004 up to 2012.7. The overall increase beyond the HTS is a factor ~10. This increase is due in part to the recovery of intensity in the 11-year solar modulation cycle (temporal) and partly due to gradients in the heliosheath (spatial).

At V2 the intensity of these GCR H nuclei at the time of the HTS crossing (2.7 years later than V1) was initially ~2.5 times larger than at the time V1 crossed the HTS (e.g., Stone, et al., 2008).  This difference is due to the later crossing of the HTS by V2 during which time the overall GCR intensity increased (temporal).  After the HTS crossing at V2, however, the intensity of this H component increased very little over the next 3 years.  Then starting in late 2010 the intensity increased by over 50% in a 1 year period.  The maximum rate of increase of this GCR H component coincides with the maximum rate of increase of GCR electrons at about 2011.25.  After about 2011.8 and up to the present time (~5 AU of outward movement of V2) there has been little intensity change in the GCR H component.

**3). TSP and ACR at V1 and V2.**  For the TSP and ACR intensity variations at V1 and V2 one needs to carefully compare Figures 3 and 4 where the "features" in each hemisphere can be directly compared relative to the overall scales of the N-S heliosheath.   Not only are the radial intensity profiles of these two components in the N and S heliosheaths greatly different, but the individual TSP and ACR intensities themselves are different in each heliosheath.  These spectral and temporal variations have been described at V1 and V2 by Cummings et al., 2007, 2010 and 2011, who have also made N-S heliosphere comparisons of the higher energy ACR. What we discuss here is a summary of weekly average rates with particular emphasis on specific N-S heliospheric comparisons throughout the extent of the heliosheath, separately for each species, TSP and ACR.

We first consider the 28-42 MeV He nuclei in Figure 3, which are mainly ACR.  At V1 in the N heliosheath the intensity of these ACR is very low at the time of the HTS crossing, and increases steadily by nearly a factor ~5 to a maximum at about 2009.5 and then decreases slowly so that at 2012.5 at V1, just before their sudden disappearance (shown here at 2015.4), the intensity is ~50% less than at the maximum in 2009.  At V2 in the S heliosheath, however, the intensity of ACR, which was initially ~3 times larger than at V1 when it crossed the HTS, does not change significantly for ~3 years and then increases by a factor ~2 so that by the end of 2012 the intensity of ACR at V2 is actually ~1.5 times larger than that measured at V1 at the time of maximum at ~2010.0 in the N heliosheath.

For the TSP shown in Figure 4 the intensity differences in the two hemispheres are large and complex and different from those of ACR. At V1 the largest intensity peak for the TSP actually occurs at ~3.5 AU beyond the HTS (at 2005.9). This peak, which is a factor of ~2 times higher than the intensity maximum of TSP observed at V1 later on at 2010.0 when it is 16 AU beyond the HTS. This "feature" has been earlier described by Webber, et al., 2009, as being due to an event, that looks much like "a solar cosmic ray event observed at the Earth" e.g., a predominately low energy event where the higher energy particles arrive first. This is significant because later at V2, the increase at ~2009.15 which is shown in Figure 4 and described in the next section has many similar characteristics and the same peak intensity as the event occurring at 2005.9 at V1.

After about 2008 the TSP intensities at V1 are much smoother and slowly varying with little change until late in 2011. At this time, at a radial distance ~118 AU, there is a sudden increase followed by an irregular decline in these particles which ended with the disappearance of the TSP (along with ACR) starting at 2012.65 (121.7 AU).

At V2 a large increase in TSP occurs just after the HTS crossing and this timing has been discussed in a separate paper (Webber, et al., 2012). We believe that this increase and the associated 42 and 21-day periodic variations of TSP during this time period, which are, in fact, out of phase with electrons, may indicate the local acceleration of both TSP and electrons during this time period. These increases may be related to the arrival of a large IP shock at the HP. The peak intensities in the 1.9-2.7 MeV TSP H channel at V2 at this time were ~2 times higher than the peak intensities seen at any time at V1.

The 2$^{nd}$ large increase in TSP at V2 that peaks at 2009.15 when V2 is ~4-5 AU beyond the HTS is discussed in the next section. It has a peak intensity at 1.9-2.7 MeV H which is comparable to the peak intensity of these same particles seen at V1 at 2005.9 when it was ~3.5 AU beyond the HTS.

Outside of the two large increases at V2 just noted, the TSP intensity at all energies changed very little beyond the HTS and actually overall has decreased slightly up to the present time.

**The Event #2 at V2, With an Intensity Peak of TSP at 2009.15 and an Extended Period of Intensity Periodicities with an Average Period of ~27 Days**

This discussion covers the time interval leading up to the event #2 at V2 with the peak intensity occurring at 2009.15, as well as the interval after the event when periodic intensity variations with an average peak to peak time of 27 days, lasting for several months were observed. This covers the overall time interval from 2008.7 to 2010.3 when V2 was between 3-8 AU beyond the HTS. The data at several TSP energies, for GCR electrons and also for the higher rigidity 27-42 MeV He ACR, during this event, are shown in Figure 5.

Prior to the start of the rapid TSP intensity increase in this event which started at 2009.05 at all energies, strong intensity peaks of TSP were observed earlier at 2008.71, 2008.87 and 2009.01, with an average periodicity ~52 ±5 days. Much weaker electron intensity peaks were also observed at this time, out of phase with the TSP peaks. Also just prior to the >0.5 MeV intensity peak which occurred at 2009.16, the intensity of 6-12 MeV H nuclei reached a peak at 2009.10, followed by the 3-8 MeV H nuclei peak at 2009.12, then the 1.9-2.7 MeV H nuclei peak at 2009.14 and lastly the >0.5 MeV peak. This energy dependence of the peak intensity is typical of the onset of a solar flare event at the Earth where the times of the intensity peaks at different energies are related to the interplanetary diffusion between the Sun and the Earth.

After the intensity peak at 2009.16 for 0.5 MeV particles, the intensities at all TSP energies decay rapidly so that by 2009.4 the TSP intensities at all energies are less than they were before the event began at 2008.5. After 2009.4, begins an interval of nearly a year when a periodic variation of average ~27 days but with considerable variability is observed, mainly for TSP. The vertical lines in Figure 5 indicate the times of the intensity peaks during this time interval. The numbers between the lines indicate the time in days between peaks.

We conclude from this discussion that between about 2009.0 and 2009.4 V2 passed into a region where the nominal 27 day co-rotating structure of the current sheet was prominent. As V2 passed into this structure the intensities of higher energy GCR H and He nuclei decreased slightly (see Figure 2) much like a "Forbush decrease" in cosmic ray intensity at the Earth, but on a much larger scale in the heliosheath. Also this passage is associated with the observed increase (and possible acceleration) of both TSP and electrons (see Figures 1 and 4) much like what occurs near an interplanetary shock inside the HTS, such as the well known Halloween event shock that was observed by V2 at 73 AU (Richardson, et al., 2005; Intriligator, et al., 2008; Webber, Intriligator and Decker, 2012).

It should be noted that the time of peak TSP intensity near 2009.15 coincides with an abrupt change in direction of the magnetic field at V2 as reported by Burlaga, et al., 2010. The magnetic field changes from ~270° to 90° at about 2009.15 and remains at ~90° until about 2009.18 when it again returns to 270°. This is the longest of several of these sudden short field direction reversals at V2 seen by Burlaga, et al., 2010, during the time period from 2007.6 to 2009.4. Other prominent field direction reversals occurred much earlier at 2007.97 and 2008.20, in conjunction with event #1 at V2. Both of these direction reversals were also accompanied by striking TSP and electron intensity changes.

The TSP intensity variations with an average period 27 days after 2009.4 continued until early 2010. The amplitude and period of these variations were variable. The time interval from 2009.6-2010.3 at V2 when the 27 day periodicities were observed has been described by Burlaga, et al., 2011, as an unipolar zone where the magnetic field is directed primarily away from the Sun (~90°). Our data thus supports the argument by Burlaga, et al., 2011, that V2 may have crossed the boundary of the maximum latitudinal extent of the HCS sometime between about 2009.0 and 2009.4 (see also later discussion).

**The Large Increase of Electrons and ACR at V2 with a Maximum Rate of Increase at 2011.25 at a Distance ~11 AU Beyond the HTS Crossing (Event #3 at V2)**

Expressed as pct. /AU, the change of 5-12 MeV electrons in the time period when the maximum intensity change was observed between 2011.22-2011.30 (which is ~27 days in duration) is ~185%/AU, the largest relative increase in electrons observed at V1 or V2 over a period of 1 AU. Further details of this event at V2 will be discussed in the next section after we have introduced the concept of a wavy heliospheric current sheet (HCS).

**Heliosheath Intensity Differences Observed by V1 and V2 and the Role of the HCS and Other Plasma and Field Parameters**

Prior to the extended measurements of the S heliosheath by V2 that are reported here and have been described elsewhere (Cummings, et al., 2011, 2012; McDonald, et al., 2012), the initial anticipation was that the S heliosheath would be similar to that in the N as observed by V1, perhaps squashed in thickness by a similar factor of 1.125 to that found for the N-S HTS crossing distances (Stone, et al., 2008; Opher, et al., 2009; Webber and Intriligator, 2011). The observations at V2 illustrate a quite different S heliosheath, however, much more structured and time varying.  At least 3 specific structures and/or features have already been passed by V2 in the 1st 12-15 AU beyond the HTS in the S heliosheath.  V1 did not observe its 1st major structure until it was 16.6 AU beyond the HTS location in the N heliosheath.  There is, however, an earlier feature observed at V1 at about 2005.9 when this spacecraft was only 3.5 AU beyond the HTS.  This large increase in low energy TSP at V1 has been described by Webber, et al., 2009, as looking much like "a solar cosmic ray event observed at the Earth", e.g., a predominately low energy event where the higher energy particles arrive first.  This is also exactly how we have described above the event #2 at V2 which had intensity peaks that occurred at 2009.15 when V2 was ~4.5 AU beyond the HTS.  The peak intensities for 1.9-2.7 MeV TSP nuclei were almost identical at the two times and locations of V1 and V2 in the N-S heliosheaths for these events which we believe may be "matching" features in the N-S heliosheath.

To examine the possible origins of these features and other features/structures observed at both V1 and V2 more closely we first consider the maximum N-S tilt angle of the HCS at the Sun in both hemispheres as derived by the Wilcox Solar Observatory (WSO) in Figure 6.  Figure 6 is an average of two techniques for determining this maximum tilt latitude, which differ typically by ± 5-10°, or more (http://wso.stanford.edu/Tilts.html).  The data has been delayed by 270 days to compare with the solar wind travel time to the HTS.  This time of 270 days is obtained by using the calculations of Zank, et al., 2001, which show that a strong shock will take an average of 270 days to reach an HTS at 90 AU.  A similar approach has been used for the S hemisphere by Richardson and Wang, 2011, and by Hill, et al., 2014, except they both use the largest tilt values instead of the average and thus get a 5-10° or larger maximum tilt angle for the HCS.    This is a major difference between these papers.

It is seen from Figure 6 that before about 2006 when V1 is at ~32° N, V1 is still within the HCS latitude region (shaded), 1st passing above it at about 2005.9 as the HCS maximum latitude decreased with decreasing solar activity (see also Burlaga, et al., 2007).   V1 then remained above the maximum HCS latitude for at least 4 years until 2010 when this maximum latitude began to increase due to increasing solar activity and the possible effects of fold back of the HCS as Voyager proceeded into the heliosheath (Pogorelov, et al., 2009).  This scenario could explain the smooth nature of the V1 intensity-time curves in the heliosheath during the time period from about 2006.0 to 2010 and also for the 1st feature at V1 mentioned earlier which occurred at about 2005.9, which would then be the result of V1 beginning to pass outward through this boundary. This time is shown by the 2[nd] arrow along the V1 trajectory in Figure 6.

For V2, on the other hand, it appears from Figure 6 that the spacecraft, which is initially at a latitude ~26°, would have crossed the HTS and entered the heliosheath at 2007.66 at a latitude less than the maximum extent of the HCS, in other words still within the sector region. The interplanetary shock from the December, 2006 events on the Sun would thus have a strong effect when it arrived at V2 late in 2007. This time is shown by the 1[st] lower arrow in Figure 6.

Sometime after 2007, however, the maximum extent of the HCS, which is slowly getting smaller with time as a result of decreasing solar activity, could have passed below the latitude of V2. We believe that this could have been at the time of the event #2 at about 2009.15 at V2 as per our earlier discussion. This time is shown by a second lower arrow in Figure 6. In this case, this event at V2 would be an analogous event to the feature observed at V1 at 2005.9, e.g., both spacecraft passed through the boundary of the sector region from below to above the maximum sector related latitudes and both saw similar intensity-time profiles of TSP.

Other factors are certainly involved in estimating the maximum latitude of the HCS influence, however. Perhaps the most important of these is the fold-back of the entire HCS structure near the outer boundary of the heliosheath as described by Florinski, 2011 and Pogorelov, et al., 2009. This aspect is perhaps more relevant in the N-hemisphere where V1 is further beyond the HTS and closer to the outer boundary of the heliosheath. This fold-back, which also carries the HCS structure from the previous solar 11-year cycles back toward the tail of the heliosphere along the flanks of the heliosheath, was described earlier by Nearny, Suess and Schmakl, 1995. Calculations by Pogorelov, Borovikov, Zank and Ogino, 2009, and Florinski, et al., 2011 show that this fold back is equivalent to an increase in the latitude of maximum extent of the sector region as V1 (or V2) gets closer to the outer boundary. In fact, if the spacecraft latitude is initially high enough, as is the case for V1 in 2011, the spacecraft can pass back through this HCS maximum extent boundary and then later pass through several structures in this folded back region for the current 11-year solar cycle before finally encountering the field from the previous 11-year cycle. This is essentially the explanation that Florinski, et al., 2011 used to explain the two sudden increases in electron intensities and change of electron gradient observed by V1 at 2009.7 and about 1.5 years later at 2011.2 at about 16.6 and 22.4 AU beyond the HTS crossing distance (see Webber, et al., 2012) as V1 then re-entered the folded back region of the HCS (arrows 3 and 4 along the V1 trajectory).

Consider now the event (a large increase in ACR) at V2 that also occurred at about 2011.2, about the same time as event #4 at V1 just discussed. The maximum extent of the HCS was rapidly increasing due to solar activity at this time. So V2 could have re-crossed the HCS boundary that it originally crossed from inside to outside in association with event #2 at 2009.15. Thus after 2011.2 V2 is therefore inside the sector structure. This is indicated by the 3[rd] lower arrow in Figure 6 at about 2011.2.

In Figure 7 we show the magnetic pressure data for V1 in the N heliosheath (there is no plasma data). Also shown are both the plasma dynamic pressure and B field pressure at V2 in the S heliosheath (lower curves) during the time period from 2008 to 2013. This is compiled from data at the COHO data base (http://cohoweb.gsfc.nasa.gov/) and recent papers by Burlaga and Ness, 2011, 2012 and Burlaga, Ness and Stone, 2013. At V1 in the N heliosphere a B field direction change ~180°, from 90° to 270°, followed by a sharp spike in B field amplitude (max ~0.3 nT), shown as arrow 3, was coincident with the 1[st] electron intensity jump at 2009.7 seen in Figure 1 (Webber, et al., 2012; Burlaga and Ness, 2010). This is the 1[st] sustained period of 270° polarity in the N hemisphere and lasts ~100 days. The 2[nd] electron intensity jump at 2011.3 at V1 was in conjunction with the first of several large spikes in B field amplitude and is shown by the upper arrow 4 in Figure 7. The B field magnitudes at the peak of the spikes at 2011.3, 2011.8 and a still later spike at 2012.35 are all larger than the largest magnitudes seen earlier at V1 just before and just after the HTS crossing. At the time of these B field spikes at 2011.3 and later, intensity changes were observed in the higher energy GCR component as well, although these changes were not as large as those for electrons. In fact the changes of these GCR and the lower energy TSP were generally in the opposite direction at these specific times.

Now turning to V2, the large increase of cosmic ray electrons and ACR also at about 2011.2 is notable in the plasma pressure data shown at the bottom of Figure 7 (lower arrow #3 in Figures 6 and 7). At about 2011.3 the plasma pressure began an increase of about 50% along with ensuing maximum intensity cycles with a periodicity of about 180 days. Note that earlier we suggested that, from the HCS data separately, that V2 passed below the maximum latitude of the sector zone at a latitude ~33° at about 2011.3.

The V2 energetic particle data presented in this paper shows no evidence yet, however, of the features at V1 seen in the particle data or the magnetic field data in the outermost 7 AU (2 years) of the N heliosheath. All of these components will now be available for study at V2 and the magnetic field (not available beyond 2010 at present) will be a crucial aspect. Will the large increases in B observed just prior to the final transition in the N heliosphere by V1 also be observed at V2 and if so what will the changes look like?

**The Thickness of the N-S Heliosheaths**

The recent passage of V1 through a distinct "trapping" boundary for TSP and ACR on August 25[th], 2012 (Webber and McDonald, 2013; Stone, et al., 2013; Krimigis, et al., 2013) now recognized as the heliopause (Gurnett, et al., 2013) has been a major step in understanding the radial extent of the N heliosheath. In particular the radial distance of this feature at 121.7 AU, leads to a N heliosheath extent of 27.6 AU corresponding to the HTS crossing distance of 94.1 AU, which occurred at the end of 2004. This feature at 121.7 AU is close to the distance of ~120 AU predicted earlier for the heliopause by Webber and Intriligator, 2011.

At the times of both of the N and S HTS crossings at 94.1 and 83.7 AU respectively, the average solar wind pressure as deduced from the normalized HTS distances calculated in Webber and Intriligator, 2011, is the same at both crossings within ± 10%. Using this uncertainty then much of the N/S ratio of 1.125 in the HTS crossing distances could be accounted for by solar wind pressure differences. If we assume that the entire factor of 1.125 that applies to the ratio of N-S HTS distances applies to the heliosheath thicknesses as well, a nominal S heliosheath thickness of 24.5 ± 2.5 AU is obtained. V2 will reach a distance of 83.7+ (24.5 ± 2.5) AU = (108.2±2.5) AU between about the middle of 2014 to and the end of 2015.

These simple calculations do not take into account the possible relative motion of the outer heliosheath boundary itself, which according to Washimi, et al., 2011, is likely to be small.

**Summary and Conclusions**

Differences in the intensity-time profiles of TSP, ACR and GCR nuclei and electrons in the N-S heliosheath are examined in this paper. In general the radial intensity profiles of all particles are smoother and more regular in the N heliosheath as observed by V1. For both lower energy TSP and higher energy ACR the intensities at V1 in the N heliosheath appear to steadily increase and to reach a maximum at about 2009.5, ~16 AU beyond the HTS crossing distance.

For GCR electrons and GCR nuclei the intensities in the N heliosheath also increase beyond the HTS crossing distance but with no evidence of a maximum at about 2009.5. In the case of electrons up to 2012.65 when V1 crossed the heliopause, this total increase is a factor of over 100. The most prominent individual features at V1 in the N hemisphere are; 1) the sudden increase of TSP and electrons at about 2005.9 when V1 was ~3.5 AU beyond the HTS crossing distance and 2) the two sudden increases in the GCR electron intensity at 2009.7 and 2011.3 (=16.6 and +22.4 AU beyond the HTS). These electron intensity increases at V1 are accompanied by large changes in the apparent electron radial intensity gradient as well as magnetic field intensity and polarity changes. They have been discussed in Webber, et al., 2012, and Burlaga and Ness, 2010, and interpreted in terms of magnetic structure features related to the fold-back of the HCS by Florinski, 2011. Their times are shown by 3[rd] and 4[th] arrows in Figures 6 and 7.

Later on at V1 at 2012.35 and 2012.65 both GCR electrons and nuclei increased suddenly. The total increase of these GCR at these times was a factor ~2 for electrons and nearly 50% for nuclei above ~150 MeV. The TSP and ACR intensities behaved differently at these times. Neither of the TSP or ACR components, which are accelerated within the heliosheath, changed significantly at 2012.35 when the GCR increased but when the GCR increased again at 2012.65 both TSP and ACR completely disappeared. These intensity changes are evidence of further large magnetic structures in the outermost heliosheath that need to be interpreted in future papers.

Turning now to V2 in the S heliosheath, the intensity variations, particularly for TSP and electrons, are complicated, starting with large intensity variations with an intensity-time period of 42 days, mainly of TSP but also including electrons, which were observed just after the HTS crossing. This is event #1 at V2. The intensity peaks during this time were ~21 days out of phase between the nuclei and the electrons. The most prominent of these intensity changes of TSP and electrons occurred at times of changes ~180° in the magnetic field direction. This period, just after the HTS crossing, is also a time of significant spectral changes in the 1-10 MeV TSP energy range (e.g., Cummings, et al., 2009). This event #1 is described in more detail in Webber and Intriligator, 2011, where it is related to the arrival at the HTS of the large IP shock from the December 2006 events on the Sun and is believed to be a time when significant acceleration of both TSP and electrons occurs. The TSP intensities at a few MeV at this time were the highest seen at either V1 or V2 in the heliosheath.

These strong periodicities in TSP intensities at V2 that were observed after the HTS crossing continued for over a year with the period increasing to 52 days until a second large transient intensity increase, again mainly in TSP intensities, occurred early in 2009 when V2 was now 4-5 AU beyond the HTS. The peak time of this intensity increase was energy dependent with the higher energies peaking $1^{st}$. This is event #2 at V2. This event also produced <u>decreases</u> of >135 MeV GCR H and He at V2 at the same time (see Figure 2), much like a Forbush decrease at the Earth. After the peak intensity of these TSP at about 2009.15, the intensities decreased rapidly and a prominent periodic intensity variation with an average peak to peak period of 27 days was observed which lasted for several months. After that time period, in early 2010, when V2 was now ~8 AU beyond the HTS crossing location, the intensities of all components, GCR nuclei and electrons as well as TSP and ACR, were still roughly the same as they were just <u>before</u> the HTS crossing itself, 2.5 years earlier.

We believe that the event #2 at V2 in early 2009, after which the 26-day periodicities were first observed, may indicate the passage of V2 from inside the HCS zone of latitudes at ~30° S to a region just outside the boundary of this zone. This is as a result of the decreasing maximum latitude of the HCS at the Sun at this time. This time is shown by the $2^{nd}$ lower arrow in Figures 6 and 7.

After about 2011.0 at V2, a remarkable increase of all components, but particularly electrons and higher energy ACR, began with the maximum rate of increase occurring at about 2011.25, when V2 was 11.1 AU beyond the HTS. This is event #3 at V2 shown by arrow 3 in Figures 6 and 7. After the maximum intensities at 2011.3 the intensities of all components then remained nearly constant for several months and began to decrease slowly and irregularly. This decrease continued to the end of 2012 when V2 was ~17.0 AU beyond the HTS crossing distance. The overall increase for electrons during the onset of this event from 2010.9 to 2011.3 was a factor ~10.

Event #3 at V2, which occurred about 2.1 years after event #2, could thus be the result of this spacecraft passing back again into the HCS sector zone. The increase in maximum latitude of the HCS could be the result of the increasing solar activity that began in early 2010, projected out to the location of V2 as well as the fold back of current sheet as it approaches the HP. This would mean that V2 has spent ~2 years (6 AU) outside the maximum extent of the HCS between events 2 and 3 (see Figure 6). This represents a time period at V2 when the solar wind pressure slowly decreased. Magnetic field data for this corresponding period, after 2010.0, are not available.

At 2014.0, V2 is at ~103.7 AU from the Sun. This is 20.0 AU beyond the S HTS crossing location of 83.7 AU or about 0.72 times the total heliosheath thickness of 27.6 AU determined in the N hemisphere. Estimates earlier in this paper using the N-S HTS crossing distance ratio of 1.125 have led to an estimated S heliosheath thickness ~24.5 AU with an uncertainty of perhaps ±2.5 AU due to possible solar wind pressure differences at the times of HTS crossing. If these estimates are correct, V2 should reach the location of the S heliosphere trapping/modulation boundary between mid-2014 and late 2015.

**Acknowledgements:** The authors wish to thank JPL for financial support of the Voyager project under the direction of project scientist Ed Stone. The data used herein come partly from data which may be accessed through the Voyager CRS public web-site at http://voyager.gsfc.nasa.gov/, partly from CRS published data and partly from internally available web-sites maintained by Nand Lal and Bryant Heikkila. The contributions of Frank McDonald and Alan Cummings are also gratefully acknowledged. We also thank the Carmel Research Center for their support of Devrie Intriligator.

# FIGURE CAPTIONS

**Figure 1:** Five day running average at V1 and V2 of 5-12 MeV electrons. The time is the time at V2. V1 times are delayed by 2.70 years. The reason for this delay is described in the text. The statistical uncertainty is about the size of the individual data points in this and in Figures 2-5. The radial distance in AU of V2 is shown on the top scale.

**Figure 2:** Identical to Figure 1 except it is for 130-242 MeV GCR H nuclei. The intensities here are weekly averages.

**Figure 3:** Identical to Figure 2 except that the intensities are weekly averages for 28-42.5 MeV/nuc He nuclei which consist mainly of ACR.

**Figure 4:** Identical to Figure 2 except that the intensities are weekly averages for 1.9-2.7 MeV H nuclei which consist mainly of TSP.

**Figure 5:** 5 day running averages of various intensities during the time interval of event #2 at V2. 0.5 MeV H in cts/s; 1.9-2.7 MeV H (x24); 27-42 MeV ACR He (x3.3) (weekly average); and 5-12 MeV e excess above background (x10$^3$). Vertical lines and numbers are used to describe periodic variations (see text).

**Figure 6:** Average of two methods to determine the maximum latitudinal extent of the HCS near the Sun in the N-S hemisphere from 2004 to 2013 (from WSO web-site, John Hoeksema, http://wso.stanford.edu/ Tilts.html). Red curve is- 3x running average of 27 day average data. Time is delayed by 270 days to compare with the average solar wind travel time to the N-S HTS. The slowly changing V1 and V2 latitudes are shown as solid lines. The shaded region is inside the estimated region of maximum HCS extent. The upper and lower arrows with #'s from 1 to 4 are at the times of significant events in the energetic particle rates at V1 and V2 as described in the text.

**Figure 7:** 27 day running averages for the B field pressure at V1 from 2004-2012 (top), and for the B field (blue) and solar wind dynamic pressure (red) at V2 from 2006 to 2013 (bottom) from http://web.mit.edu and COHO. The arrows are at the same times as in Figure 6 and are described in the text.

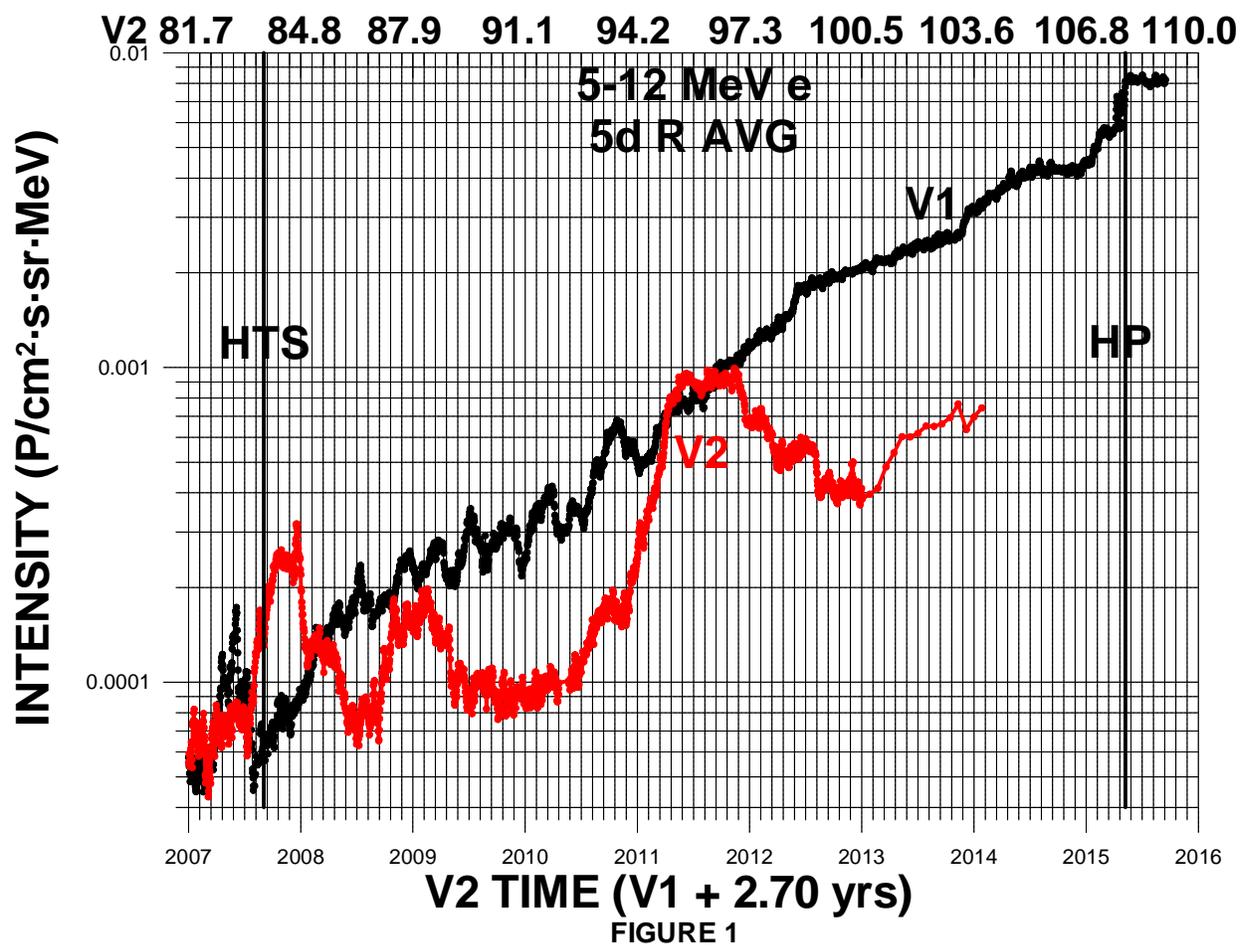

FIGURE 1

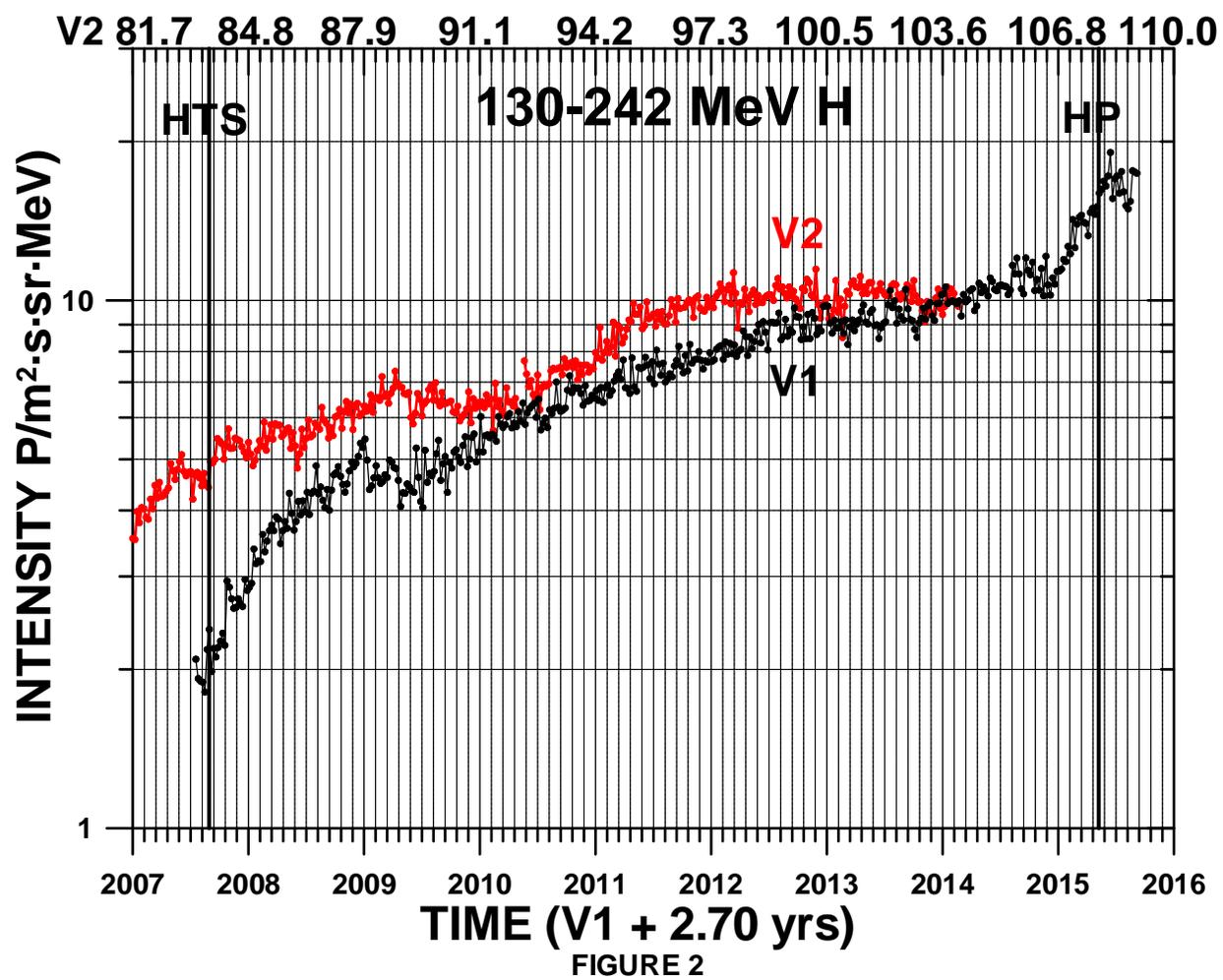

FIGURE 2

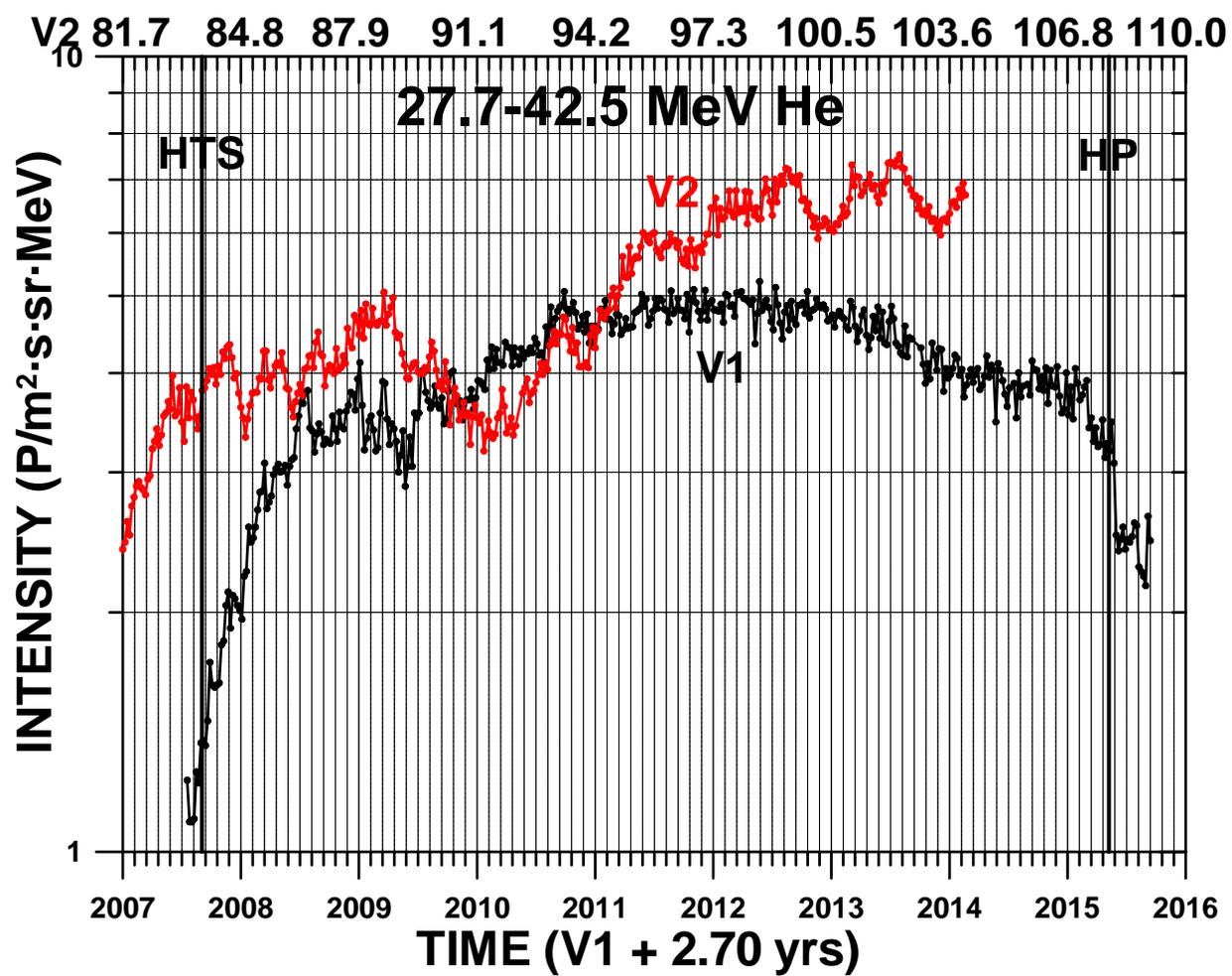

FIGURE 3

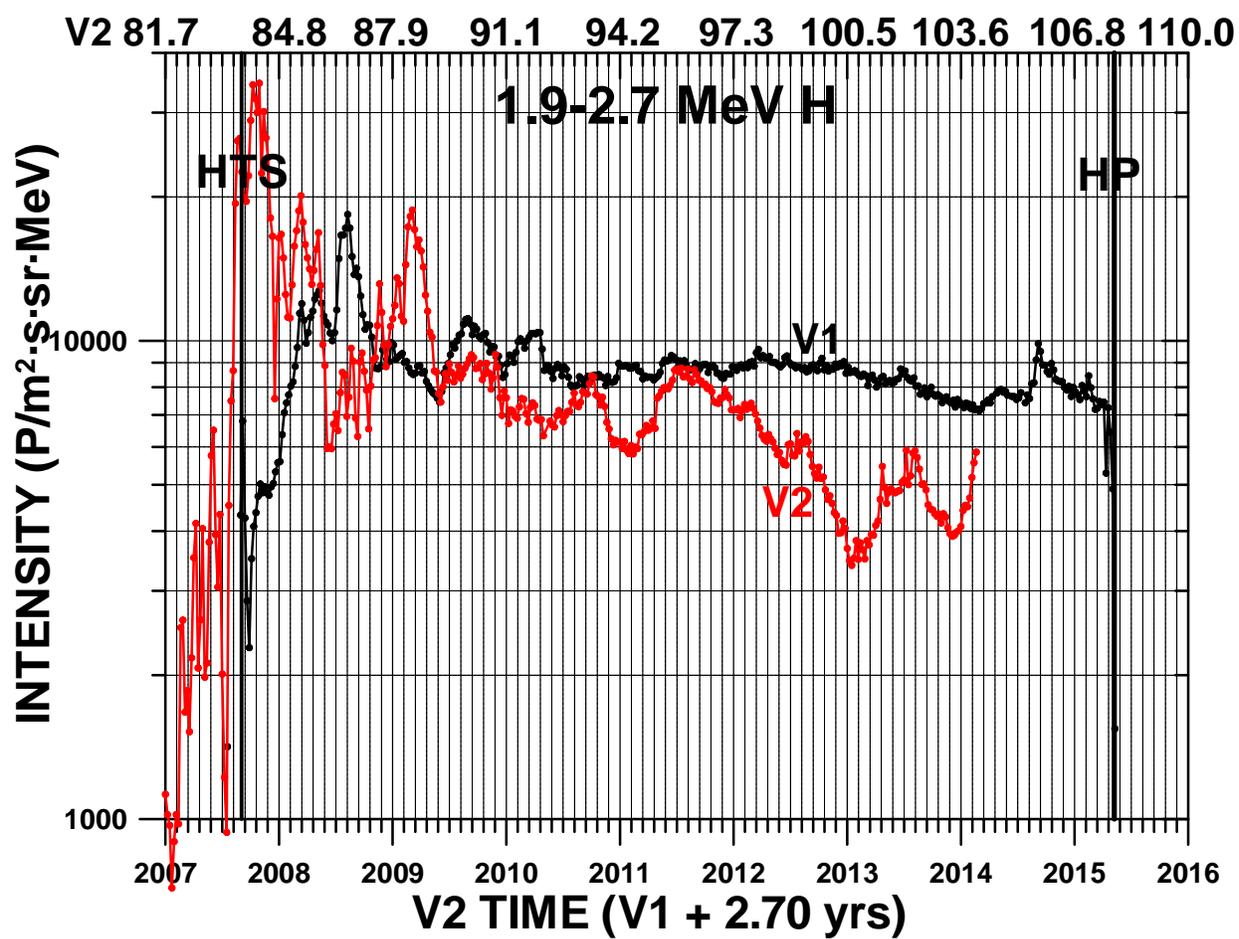

FIGURE 4

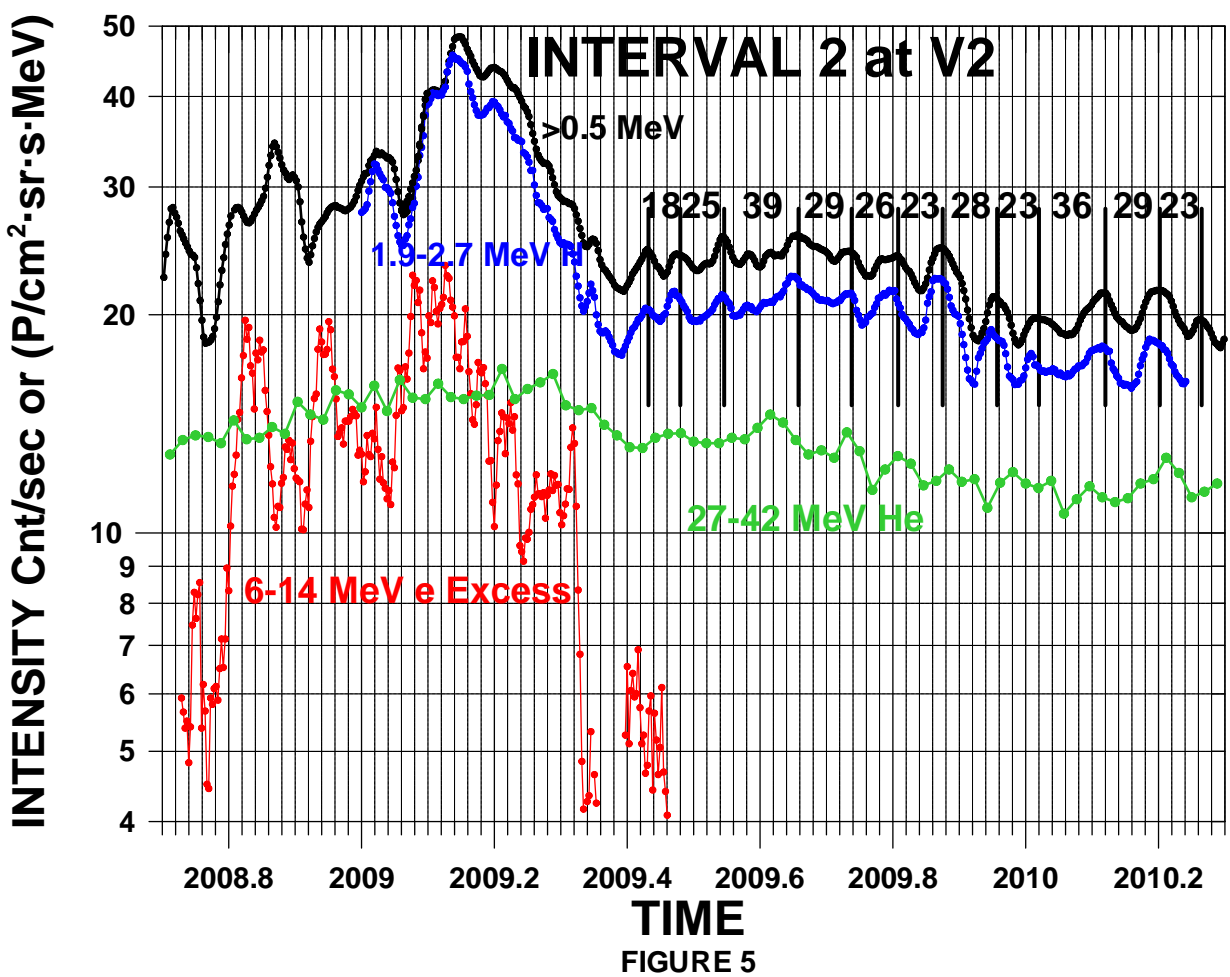

FIGURE 5

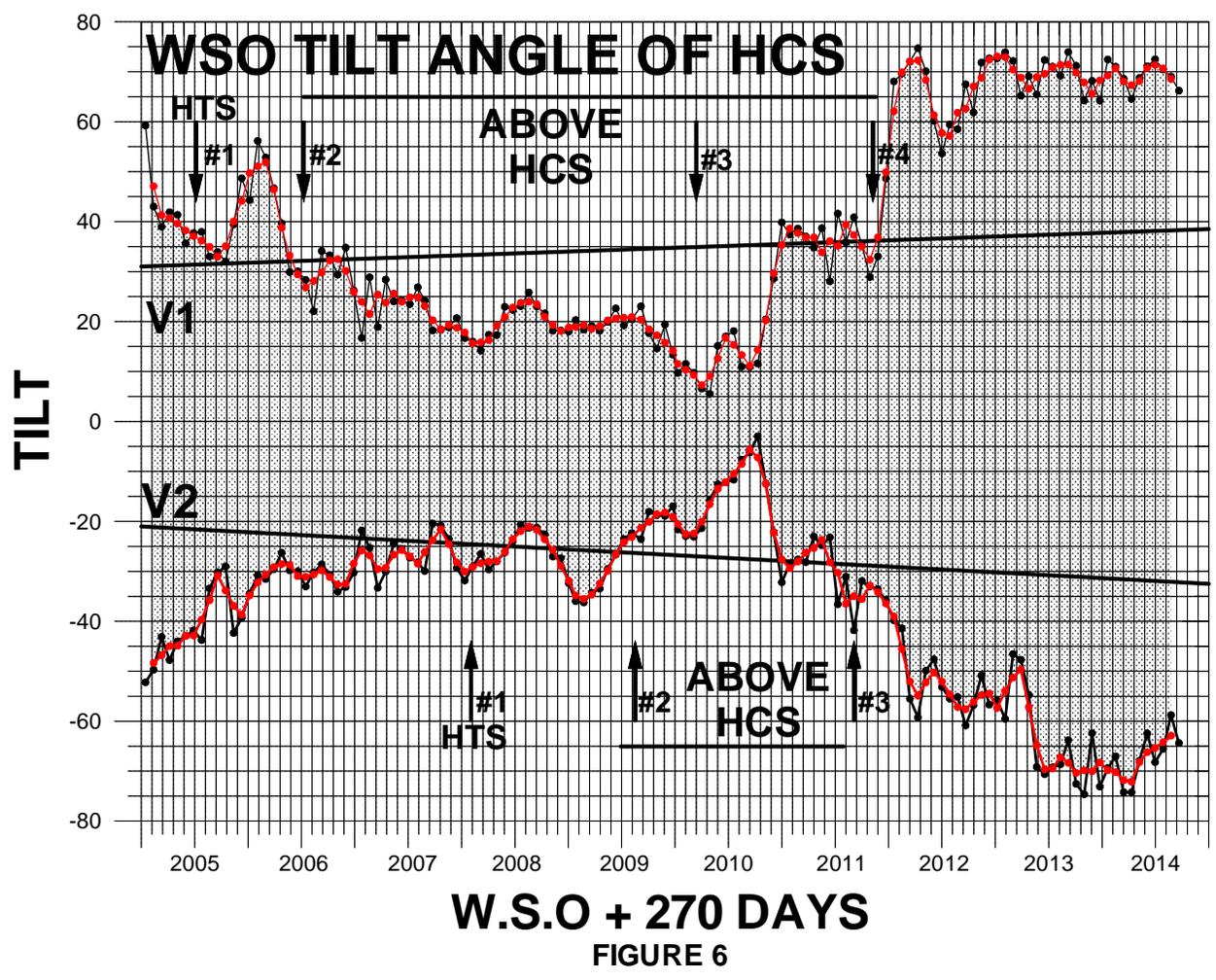

FIGURE 6

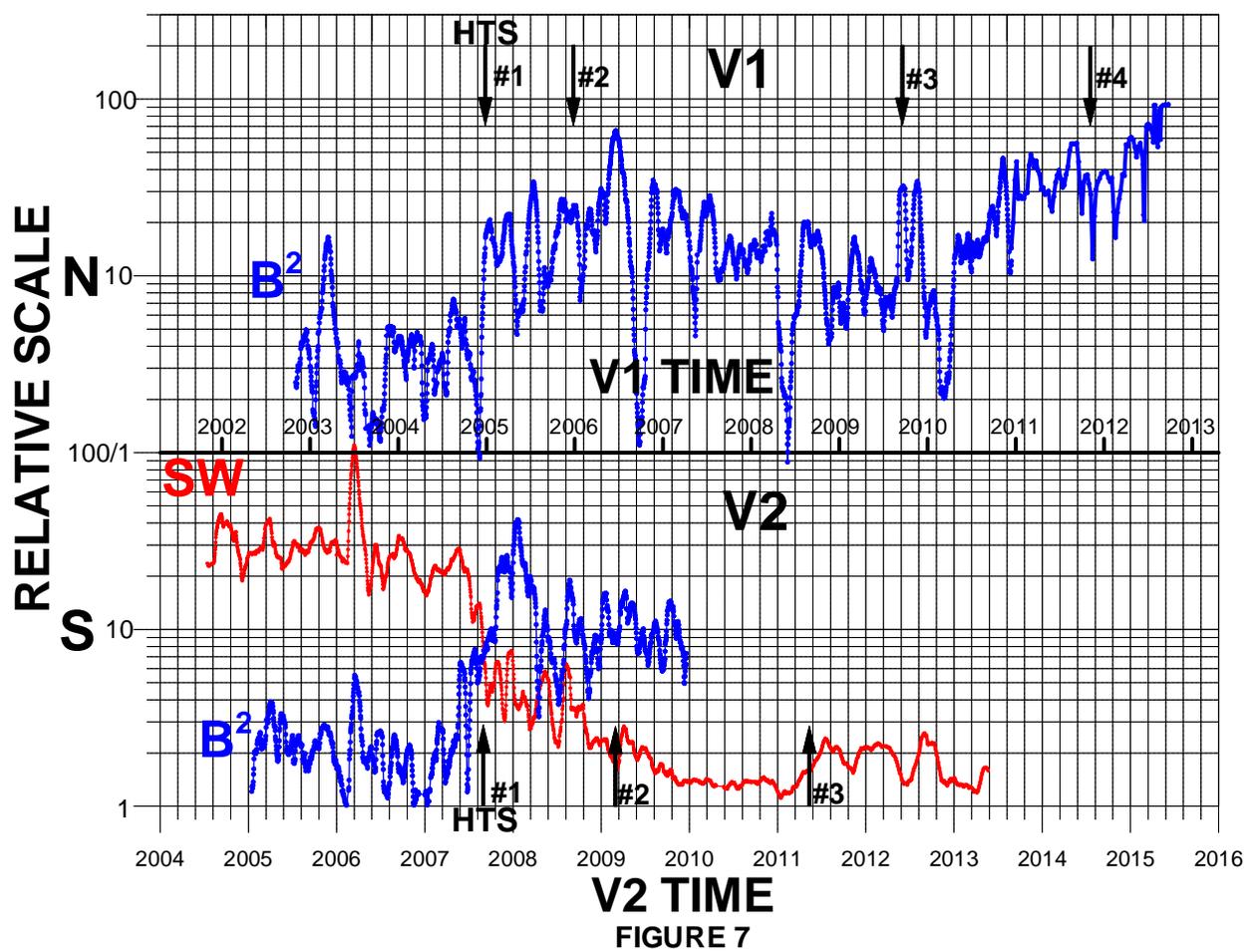

FIGURE 7